\documentclass[aps,prl,twocolumn,groupedaddress,showpacs]{revtex4}

\usepackage{graphicx}
\usepackage{amsmath}

\begin{document}

\title{The Pseudogap Ground State in High Temperature Superconductors.}

\author{J.G. Storey$^1$, J.L. Tallon$^{1,2}$, G.V.M. Williams$^{2}$}

\affiliation{$^1$School of Chemical and Physical Sciences,
Victoria University, P.O. Box 600, Wellington, New Zealand}

\affiliation{$^2$MacDiarmid Institute, Industrial Research Ltd.,
P.O. Box 31310, Lower Hutt, New Zealand}

\date{\today}

\begin{abstract}
By re-examining recently-published data from angle-resolved photoemission spectroscopy we demonstrate that, in the superconducting region of the phase diagram, the pseudogap ground state is an arc metal. This scenario is consistent with results from Raman spectroscopy, specific heat and NMR. In addition, we propose an explanation for the ``Fermi pockets'' inferred from quantum oscillations in terms of a pseudogapped bilayer Fermi surface.
\end{abstract}


\maketitle

In recent years, the results from a number of experimental techniques, including heat capacity\cite{LORAM,OURWORK3,OURWORK1,WEN}, Andreev reflection\cite{DEUTSCHER}, Raman spectroscopy\cite{LETACON}, angle-resolved photoemission spectrocopy (ARPES)\cite{TANAKA,HASHIMOTO,LEE,KONDO} and scanning tunelling microscopy (STM)\cite{BOYER}, have begun to converge in support of the presence of two distinct gaps in the low-energy electronic structure of optimal- and under-doped high-$T_c$ cuprates. These are: the pseudogap which opens in the normal state below a critical hole doping of about 0.19; and the superconducting gap which opens at the onset of superconductivity.
The pseudogap presents an extra layer of complexity that must be resolved in the quest to explain the physics of high-$T_c$ superconductors.

In ARPES measurements the pseudogap is apparent as a depletion of states near the ($\pi$,0) and (0,$\pi$) regions of the Brillouin zone. As a result, the Fermi surface seems to consist of a set of disconnected ``Fermi arcs''\cite{FERMIARCS}. Below $T_c$ the superconducting gap opens on the arcs, shrouding their character at zero temperature - the pseudogap ground state - from direct observation. In what has become an influential paper, Kanigel \textit{et al}.\cite{FERMIARCS2} reported measurements of the Fermi arc length above $T_c$ as a function of temperature for various doping levels. Their results were plotted as a function of reduced temperature $t=T/T^*$ where $T^*$ is the temperature above which pseudogap effects are no longer observed in their analysis. When plotted in this fashion, the Fermi arc length (FAL) seems to extrapolate linearly to zero at $t=0$. Thus, they suggest that the $T=0$ pseudogap state is a nodal liquid, at all doping levels. Later the same year Valla \textit{et al}.\cite{VALLA2} reported that in La$_{2-x}$Ba$_x$CuO$_4$ with $x=1/8$, where superconductivity is suppressed by ``stripes'', the Fermi surface consists of four nodal points.
If, at this doping, the ground state pseudogap is exposed then these results appear to confirm its nodal character.

Such a conclusion would have fundamentally important implications for the two scenarios envisaged for the pseudogap. One views the pseudogap as a competing order parameter while the alternative is to consider the pseudogap as a pairing gap arising from a phase incoherent pairing state or from real-space local pairs. Here is the problem: given that the pseudogap energy is usually observed to exceed the superconducting gap, a nodal groundstate pseudogap would remove all states available for superconductivity. The manifest persistence of superconductivity in underdoped cuprates effectively would eliminate the competing-order-parameter scenario. The normal-state-pairing scenario alone survives because the pairs need only become coherent or condense into the BCS state. The proposed nodal-metal ground state thus provides an important arbiter of these models.
\begin{figure}
\centering
\includegraphics[width=\linewidth,clip=true,trim=0 0 0 0]{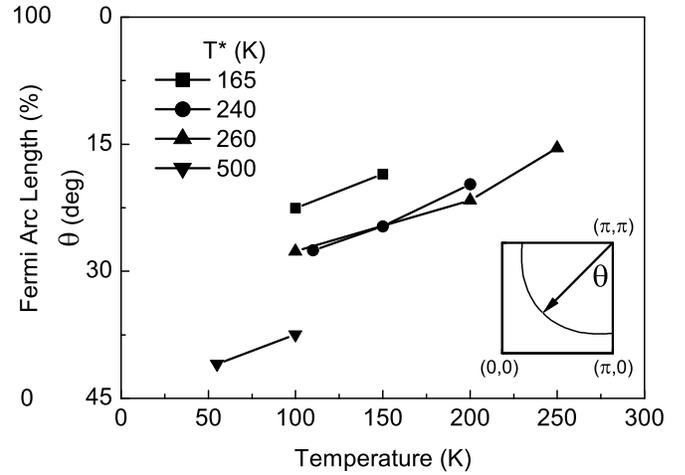}
\caption{Fermi arc length extracted from the raw data of Ref.~\cite{FERMIARCS2} plotted vs temperature. With the exception of the most underdoped sample, the Fermi arc lengths extrapolate to finite values at $T=0$. Inset: Fermi surface in the 1st quadrant of the Brillouin zone showing the angle $\theta$.}
\label{FALVST}
\end{figure}

Recently however, we were able to demonstrate that results from Raman spectroscopy paint a different picture. Raman spectroscopy provides a useful tool for studying energy gaps because the selection of different polarizations allows different regions of the Brillouin zone to be probed. The Raman data of Le Tacon \textit{et al}.\cite{LETACON} was used to test two scenarios for the pseudogap ground state: a nodal-metal state or an arc-metal state. It was found that finite FAL's at zero temperature were required to reproduce the trend observed in the nodal (B$_{2g}$) Raman response\cite{STOREYRAMAN,STOREYRAMAN2}. Two other groups reach the same conclusion (that the pseudogap ground state is an arc metal). Wen \textit{et al}.\cite{WEN} use high magnetic fields in La$_{2-x}$Sr$_x$CuO$_4$ to suppress superconductivity and they find a finite non-zero value of the specific heat coefficient at $T=0$, and Zheng \textit{et al}.\cite{ZHENG} in Bi$_2$Sr$_{2-x}$La$_x$CuO$_{6+\delta}$ find non-zero $1/T_1T$ values at $T=0$, again using high fields to expose the pseudogap ground state. Both of these indicate a finite residual density of states at $T=0$ consistent with an arc metal. In the following we seek to reconcile these apparently conflicting findings.

Because Kanigel \textit{et al}. arrive at their important conclusions by plotting the FAL vs $T/T^*$, we simply replot the data in terms of absolute temperature from the raw data shown in Figure 4a of their paper\cite{FERMIARCS2}. The resulting plot (see Fig.~\ref{FALVST}) shows that, with the exception of the most underdoped sample ($T^*=500$K), the temperature dependence of the FAL's extrapolate to non-zero values at zero temperature. Thus the conclusion of a nodal-metal ground state appears to be an artifact of the scaling analysis that conceals the important detail shown here in Fig.~\ref{FALVST}. Further, the zero temperature FAL reduces with decreasing doping, consistent with our Raman modelling where a zero-temperature FAL roughly proportional to $T_c$ was found to reproduce the data.

Here we independently determine the Fermi arc length plot from the electronic entropy as follows.
In previous work\cite{STOREYENTROPY} we calculated the electronic entropy from an ARPES-derived energy-momentum dispersion and a nodal pseudogap model based on the data from Kanigel \textit{et al}. Although excellent fits to the normal-state entropy were obtained, we found it impossible to calculate the magnitude of the superconducting gap, $\Delta(T)$, self consistently using the BCS gap equation in the presence of the pseudogap. The fully-nodal ``non-states-conserving'' pseudogap model simply removed too many states to provide a converging solution. As a result, the pseudogap was left out of the calculation of $\Delta(T)$, a somewhat artificial solution. We now note that an arc metal pseudogap model allows the pseudogap to be included in the self-consistent calculation. 
\begin{figure}
\centering
\includegraphics[width=\linewidth,clip=true,trim=0 0 0 0]{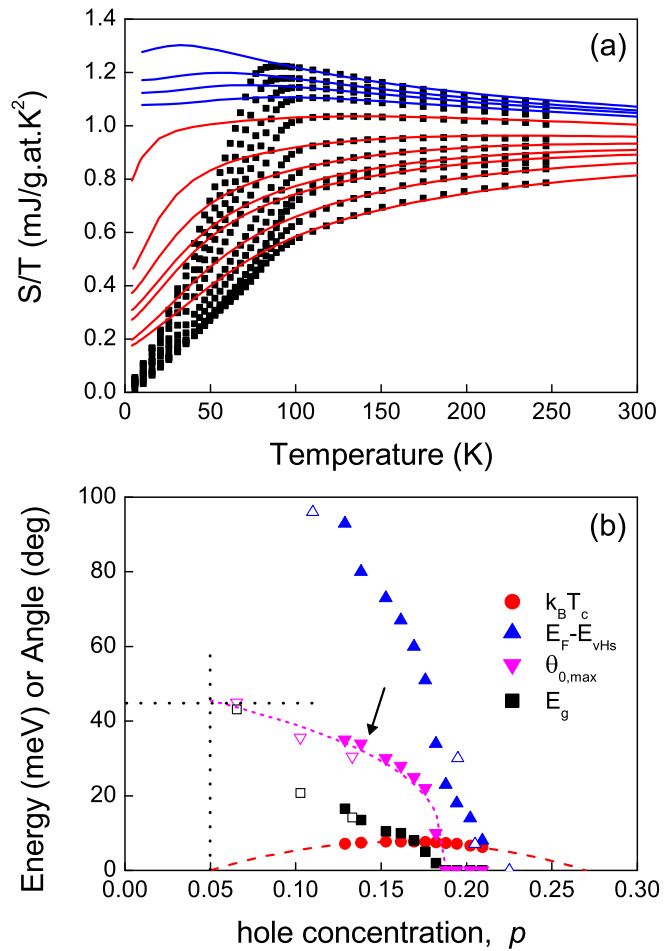}
\caption{(Color online) (a) Normal-state fits to the Bi-2212 entropy data of Loram \textit{et al}. (Ref.~\cite{ENTROPYDATA2}) assuming an arc metal pseudogap model. For clarity every 20th data point only is shown. (b) Fermi level position $E_F-E_{vHs}$, Fermi arc angle at $T=0$ $\theta_{0,max}$, and pseudogap magnitude $E_g$ extracted from the fits in (a). Open symbols are experimentally determined values: $\theta_{0,max}$ and $E_g$ is from Ref.~\cite{FERMIARCS2}, while $E_F-E_{vHs}$ is from Refs.~\cite{2212VHS,DING,UD77K}.}
\label{SVST}
\end{figure}

Shown in Fig.~\ref{SVST}(a) are normal-state fits made to the entropy data of Loram \textit{et al}.\cite{ENTROPYDATA2} using the approach described previously\cite{STOREYENTROPY}, but now with an extra parameter $\theta_{0,max}$ that defines the angle to which the pseudogap extends around the Fermi surface at zero-temperature. The temperature dependence of the FAL is now given by
\begin{equation}
\theta_0(T)=\theta_{0,max}\left[1-\tanh\left(\frac{T}{2T^*}\right)\right]
\label{THETA0EQ}
\end{equation}
The parameters extracted from the fits are shown in Fig.~\ref{SVST}(b) (solid symbols) along with experimentally determined values from the Kanigel data and other ARPES papers (open symbols). The general agreement is very good and, importantly, shows the $T=0$ Fermi arc angle (arrowed) tending towards 45 degrees as doping decreases. We consider this as a key conclusion. The location at which the Fermi arcs collapse to point nodes coincides with, and provides a natural explanation for, the disappearance of superconductivity near $p=0.05$. At this point there is no residual Fermi arc or ``spectral weight'' available for superconductivity. This in turn implies a continuity of the large-Fermi-surface concept to low doping and leads us to the issue of ``Fermi pockets'', to be discussed soon.
This conclusion, namely that the pseudogap ground state is an arc-metal and the arc smoothly contracts to the node as doping is reduced towards $p\approx0.05$, is supported in detail by the $1/T_1T$ data of Zheng \textit{et al}.\cite{ZHENG}. They find a residual non-zero value of $1/T_1T$ as $T\rightarrow0$ which is progressively reduced to zero as doping is decreased. Our analysis of their data shows a quantitatively similar doping dependence of $\theta_{0,max}$ to that shown in Fig.~\ref{SVST}(b). A similar quantitative correspondence can be found with the residual specific heat coefficient inferred by Wen and Wen\cite{WEN}.

Moreover, an exacting test of the ground state arc length is also provided by the $T$-dependence of the superfluid density $\rho_s(T)$ which, if the Fermi arc continues to shrink towards the node, would exhibit a downturn at low temperature\cite{STOREYENTROPY} that is not observed. In fact, the detailed $T$-dependence of $\rho_s(T)$ should provide strong constraints on whether, and by how much the Fermi arcs continue to shrink below $T_c$. 
The $c$-axis infra-red conductivity $\sigma_c(\omega)$ measured by Yu \textit{et al}.\cite{YU} also provides a good guide as to the evolution of the Fermi arcs below $T_c$. On cooling below $T^*$ these authors find a smooth loss of spectral weight below the pseudogap energy $\omega_{PG}$ which is transfered to higher energy above the gap. With the onset of superconductivity spectral weight below the superconducting gap is transfered to the $\omega=0$ superfluid delta function. In this way both gaps can be clearly identified. Importantly, below $T_c$ no further weight is transfered above $\omega_{PG}$ and no further loss of spectral weight occurs in the frequency range between $\omega_{SC}$ and $\omega_{PG}$. This places tight constraints on the degree to which Fermi arcs continue to shrink below $T_c$, and certainly indicates they do not collapse to the nodes.

We turn now to the issue of ``Fermi pockets''. A new development on the subject of Fermi surfaces in underdoped cuprates is the observation of quantum oscillations in the Y-123\cite{DOIRON} and Y-124\cite{YELLAND,BANGURA} systems under very high magnetic fields (up to 85T). Quantum oscillations provide a measure of the Fermi surface area via the Onsager relation, assuming that it is valid in these systems. The frequencies observed imply small nodal ``Fermi pockets'' enclosing areas of 1.9\% and 2.4\% of the Brillouin zone in YBa$_2$Cu$_3$O$_{6.5}$ and YBa$_2$Cu$_4$O$_8$ respectively. So far however, Fermi pockets have not been observed by ARPES. The two cuprate systems in which quantum oscillations have been observed to date contain double CuO$_2$ planes and are further complicated by the presence of CuO chain layers. ARPES measurements of the Fermi surfaces of these materials show that they are split into bonding and antibonding bands\cite{BORISENKO,KONDO124}, due to weak coupling between the two CuO$_2$ planes. Furthermore, this so-called bilayer splitting is large and occurs at all points around the Fermi surface, unlike in Bi-2212 where the splitting is confined mostly to the antinodes\cite{FENG}. 

\begin{figure}
\centering
\includegraphics[width=5cm,clip=true,trim=0 0 0 0]{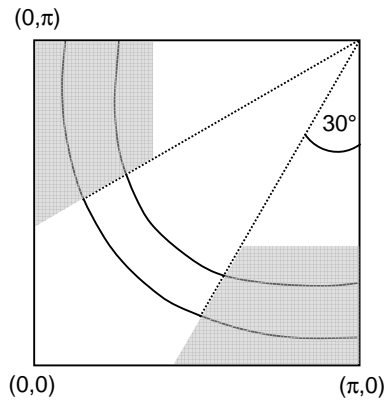}
\caption{The CuO plane Fermi surface of Y-124 determined by ARPES\cite{KONDO124} (solid lines). The area between the bonding and antibonding band Fermi arcs is approximately 1.8\% of the Brillouin zone. The Fermi surface area inferred from quantum oscillations experiments is 2.4\%.}
\label{Y124FIG}
\end{figure}

Using the above information concerning the temperature dependence of the Fermi arcs, we propose that the Fermi pockets might be explainable in terms of a pseudogapped bilayer Fermi surface with band hopping under the high magnetic field. Y-124 has an estimated hole concentration somewhere between $p=0.125$ and 0.14. Assuming that the pseudogap behaves in a similar fashion to that in Bi-2212, we estimate from Fig.~\ref{SVST}(b) that the Fermi arc extends to about 30$^\circ$ at zero temperature.
Fig.~\ref{Y124FIG} shows the bilayer Fermi surface of Y-124 measured from ARPES\cite{KONDO124} with the pseudogapped regions grayed out. The chain Fermi surface is not shown. The area between the bonding and antibonding band Fermi arcs is approximately 1.8\% of the Brillouin zone. From Luttinger's theorem the number of carriers per copper atom for four such areas is 0.144. In comparison, the area inferred from quantum oscillations is 2.4\%\cite{YELLAND,BANGURA}. The corresponding carrier concentration is 0.192, larger than the nominal hole concentration by a factor between 1.3 and 1.5. It is interesting to note that quantum oscillations measurements on Y-123 also overestimate the carrier concentration by a factor of 1.5\cite{DOIRON}. The explanation of Fermi pockets in terms of bonding and antibonding band Fermi arcs would obviously fail to explain quantum oscillations in a single layer system such as La$_{2-x}$Sr$_x$CuO$_4$. Such measurements remain to be performed.

We now return to the observations of Valla \textit{et al}.\cite{VALLA2} which reveal a nodal ground state at $p=1/8$. In the yttrium and bismuth based high-$T_c$ cuprates the isotope effect in $T_c$, $\alpha(T_c)=-(\Delta{T_c}/T_c)/(\Delta{M}/M)$, is small in overdoped samples, taking values of about 0.06. However, in the pseudogap regime it rises rapidly, exceeding 0.5 in deeply underdoped samples. Pringle \textit{et al}.\cite{PRINGLE} have described the doping dependence of the isotope effect by a model in which $\alpha(T_c)$ depends upon the ratio $E_g/\Delta_0$, where $E_g$ is the pseudogap magnitude and $\Delta_0$ is the total spectral gap. The superconducting order parameter $\Delta_{SC}=\sqrt{\Delta_0^2(\textbf{k})-E_g^2(\textbf{k})}$ is reduced by the pseudogap. In the model, a small isotope effect in the superconducting gap, $\Delta_{SC}$, becomes amplified by the presence of the pseudogap. Turning to the isotope effect data for the lanthanum based cuprates\cite{CRAWFORD2} it is found that, in addition to the rise with decreasing doping, $\alpha(T_c)$ exhibits a sharp peak at 1/8$^{th}$ doping. One possible scenario here is that at this doping, static stripes enhance the pseudogap in some fashion, thereby resulting in the complete suppression of $T_c$, and a consequent peak in the isotope effect. It follows that because stripes are unique to this doping we must treat the observed nodal pseudogap ground state as an exception rather than the rule across the phase diagram. Indeed, if the pseudogap were to be enhanced at this doping the Fermi arc length may be concurrently reduced to the nodes, just as it is by reduced doping.

In summary, we have shown that when plotted in terms of absolute temperature, recent ARPES measurements show the pseudogap ground state to be an arc metal in the superconducting region of the phase diagram. This scenario is consistent with results from Raman spectroscopy, specific heat and NMR and we therefore consider our conclusions to be robust. The length of the Fermi arcs at zero temperature decreases with decreasing doping, tending to zero near $p=0.05$. The total collapse of the Fermi arcs to point nodes at this doping provides a natural explanation for the coincident disappearance of superconductivity. Theories that attempt to reproduce the nodal scenario are probably misdirected. Lastly we have proposed an explanation for the Fermi pockets inferred from quantum oscillations measurements in terms of a pseudogapped bilayer Fermi surface. Regardless of whether a Fermi arc picture or a Fermi pocket picture turns out to be the more correct, both scenarios result in a finite density of states at the Fermi level in the range of dopings over which superconductivity occurs.


\begin{thebibliography}{31}
\expandafter\ifx\csname natexlab\endcsname\relax\def\natexlab#1{#1}\fi
\expandafter\ifx\csname bibnamefont\endcsname\relax
  \def\bibnamefont#1{#1}\fi
\expandafter\ifx\csname bibfnamefont\endcsname\relax
  \def\bibfnamefont#1{#1}\fi
\expandafter\ifx\csname citenamefont\endcsname\relax
  \def\citenamefont#1{#1}\fi
\expandafter\ifx\csname url\endcsname\relax
  \def\url#1{\texttt{#1}}\fi
\expandafter\ifx\csname urlprefix\endcsname\relax\def\urlprefix{URL }\fi
\providecommand{\bibinfo}[2]{#2}
\providecommand{\eprint}[2][]{\url{#2}}

\bibitem[{\citenamefont{Loram et~al.}(1994)\citenamefont{Loram, Mirza, Cooper,
  Liang, and Wade}}]{LORAM}
\bibinfo{author}{\bibfnamefont{J.~W.} \bibnamefont{Loram}},
  \bibinfo{author}{\bibfnamefont{K.~A.} \bibnamefont{Mirza}},
  \bibinfo{author}{\bibfnamefont{J.~R.} \bibnamefont{Cooper}},
  \bibinfo{author}{\bibfnamefont{W.~Y.} \bibnamefont{Liang}}, \bibnamefont{and}
  \bibinfo{author}{\bibfnamefont{J.~M.} \bibnamefont{Wade}},
  \bibinfo{journal}{J. Supercon.} \textbf{\bibinfo{volume}{7}},
  \bibinfo{pages}{243} (\bibinfo{year}{1994}).

\bibitem[{\citenamefont{Tallon et~al.}(1994)\citenamefont{Tallon, Willams,
  Staines, and Bernhard}}]{OURWORK3}
\bibinfo{author}{\bibfnamefont{J.~L.} \bibnamefont{Tallon}},
  \bibinfo{author}{\bibfnamefont{G.~V.~M.} \bibnamefont{Willams}},
  \bibinfo{author}{\bibfnamefont{M.~P.} \bibnamefont{Staines}},
  \bibnamefont{and} \bibinfo{author}{\bibfnamefont{C.}~\bibnamefont{Bernhard}},
  \bibinfo{journal}{Physica C} \textbf{\bibinfo{volume}{235--240}},
  \bibinfo{pages}{1821} (\bibinfo{year}{1994}).

\bibitem[{\citenamefont{Tallon and Loram}(2001)}]{OURWORK1}
\bibinfo{author}{\bibfnamefont{J.~L.} \bibnamefont{Tallon}} \bibnamefont{and}
  \bibinfo{author}{\bibfnamefont{J.~W.} \bibnamefont{Loram}},
  \bibinfo{journal}{Physica C} \textbf{\bibinfo{volume}{349}},
  \bibinfo{pages}{53} (\bibinfo{year}{2001}).

\bibitem[{\citenamefont{Wen and Wen}(2007)}]{WEN}
\bibinfo{author}{\bibfnamefont{H.~H.} \bibnamefont{Wen}} \bibnamefont{and}
  \bibinfo{author}{\bibfnamefont{X.~G.} \bibnamefont{Wen}},
  \bibinfo{journal}{Physica C} \textbf{\bibinfo{volume}{460--462}},
  \bibinfo{pages}{28} (\bibinfo{year}{2007}).

\bibitem[{\citenamefont{Deutscher}(1999)}]{DEUTSCHER}
\bibinfo{author}{\bibfnamefont{G.}~\bibnamefont{Deutscher}},
  \bibinfo{journal}{Nature} \textbf{\bibinfo{volume}{397}},
  \bibinfo{pages}{410} (\bibinfo{year}{1999}).

\bibitem[{\citenamefont{Le~Tacon et~al.}(2006)\citenamefont{Le~Tacon, Sacuto,
  Georges, Kotliar, Gallais, Colson, and Forget}}]{LETACON}
\bibinfo{author}{\bibfnamefont{M.}~\bibnamefont{Le~Tacon}},
  \bibinfo{author}{\bibfnamefont{A.}~\bibnamefont{Sacuto}},
  \bibinfo{author}{\bibfnamefont{A.}~\bibnamefont{Georges}},
  \bibinfo{author}{\bibfnamefont{G.}~\bibnamefont{Kotliar}},
  \bibinfo{author}{\bibfnamefont{Y.}~\bibnamefont{Gallais}},
  \bibinfo{author}{\bibfnamefont{D.}~\bibnamefont{Colson}}, \bibnamefont{and}
  \bibinfo{author}{\bibfnamefont{A.}~\bibnamefont{Forget}},
  \bibinfo{journal}{Nature Physics} \textbf{\bibinfo{volume}{2}},
  \bibinfo{pages}{537} (\bibinfo{year}{2006}).

\bibitem[{\citenamefont{Tanaka et~al.}(2006)\citenamefont{Tanaka, Lee, Lu,
  Fujimori, Fujii, Risdiana, Terasaki, Scalapino, Devereaux, Hussain
  et~al.}}]{TANAKA}
\bibinfo{author}{\bibfnamefont{K.}~\bibnamefont{Tanaka}},
  \bibinfo{author}{\bibfnamefont{W.~S.} \bibnamefont{Lee}},
  \bibinfo{author}{\bibfnamefont{D.~H.} \bibnamefont{Lu}},
  \bibinfo{author}{\bibfnamefont{A.}~\bibnamefont{Fujimori}},
  \bibinfo{author}{\bibfnamefont{T.}~\bibnamefont{Fujii}},
  \bibinfo{author}{\bibnamefont{Risdiana}},
  \bibinfo{author}{\bibfnamefont{I.}~\bibnamefont{Terasaki}},
  \bibinfo{author}{\bibfnamefont{D.~J.} \bibnamefont{Scalapino}},
  \bibinfo{author}{\bibfnamefont{T.~P.} \bibnamefont{Devereaux}},
  \bibinfo{author}{\bibfnamefont{Z.}~\bibnamefont{Hussain}},
  \bibnamefont{et~al.}, \bibinfo{journal}{Science}
  \textbf{\bibinfo{volume}{314}}, \bibinfo{pages}{1910} (\bibinfo{year}{2006}).

\bibitem[{\citenamefont{Hashimoto et~al.}(2007)\citenamefont{Hashimoto, Tanaka,
  Yoshida, Fujimori, Okusawa, Wakimoto, Yamada, Kakeshita, Eisaki, and
  Uchida}}]{HASHIMOTO}
\bibinfo{author}{\bibfnamefont{M.}~\bibnamefont{Hashimoto}},
  \bibinfo{author}{\bibfnamefont{K.}~\bibnamefont{Tanaka}},
  \bibinfo{author}{\bibfnamefont{T.}~\bibnamefont{Yoshida}},
  \bibinfo{author}{\bibfnamefont{A.}~\bibnamefont{Fujimori}},
  \bibinfo{author}{\bibfnamefont{M.}~\bibnamefont{Okusawa}},
  \bibinfo{author}{\bibfnamefont{S.}~\bibnamefont{Wakimoto}},
  \bibinfo{author}{\bibfnamefont{K.}~\bibnamefont{Yamada}},
  \bibinfo{author}{\bibfnamefont{T.}~\bibnamefont{Kakeshita}},
  \bibinfo{author}{\bibfnamefont{H.}~\bibnamefont{Eisaki}}, \bibnamefont{and}
  \bibinfo{author}{\bibfnamefont{S.}~\bibnamefont{Uchida}},
  \bibinfo{journal}{Physica C} \textbf{\bibinfo{volume}{460--462}},
  \bibinfo{pages}{884} (\bibinfo{year}{2007}).

\bibitem[{\citenamefont{Lee et~al.}(2007)\citenamefont{Lee, Vishik, Tanaka, Lu,
  Sasagawa, Nagaosa, Devereaux, Hussain, and Shen}}]{LEE}
\bibinfo{author}{\bibfnamefont{W.~S.} \bibnamefont{Lee}},
  \bibinfo{author}{\bibfnamefont{I.~M.} \bibnamefont{Vishik}},
  \bibinfo{author}{\bibfnamefont{K.}~\bibnamefont{Tanaka}},
  \bibinfo{author}{\bibfnamefont{D.~H.} \bibnamefont{Lu}},
  \bibinfo{author}{\bibfnamefont{T.}~\bibnamefont{Sasagawa}},
  \bibinfo{author}{\bibfnamefont{N.}~\bibnamefont{Nagaosa}},
  \bibinfo{author}{\bibfnamefont{T.~P.} \bibnamefont{Devereaux}},
  \bibinfo{author}{\bibfnamefont{Z.}~\bibnamefont{Hussain}}, \bibnamefont{and}
  \bibinfo{author}{\bibfnamefont{Z.~X.} \bibnamefont{Shen}},
  \bibinfo{journal}{Nature} \textbf{\bibinfo{volume}{450}}, \bibinfo{pages}{81}
  (\bibinfo{year}{2007}).

\bibitem[{\citenamefont{Kondo et~al.}(2007{\natexlab{a}})\citenamefont{Kondo,
  Takeuchi, Kaminski, Tsuda, and Shin}}]{KONDO}
\bibinfo{author}{\bibfnamefont{T.}~\bibnamefont{Kondo}},
  \bibinfo{author}{\bibfnamefont{T.}~\bibnamefont{Takeuchi}},
  \bibinfo{author}{\bibfnamefont{A.}~\bibnamefont{Kaminski}},
  \bibinfo{author}{\bibfnamefont{S.}~\bibnamefont{Tsuda}}, \bibnamefont{and}
  \bibinfo{author}{\bibfnamefont{S.}~\bibnamefont{Shin}},
  \bibinfo{journal}{Phys. Rev. Lett.} \textbf{\bibinfo{volume}{98}},
  \bibinfo{pages}{267004} (\bibinfo{year}{2007}{\natexlab{a}}).

\bibitem[{\citenamefont{Boyer et~al.}(2007)\citenamefont{Boyer, Wise,
  Chatterjee, Yi, Kondo, Takeuchi, Ikuta, and Hudson}}]{BOYER}
\bibinfo{author}{\bibfnamefont{M.~C.} \bibnamefont{Boyer}},
  \bibinfo{author}{\bibfnamefont{W.~D.} \bibnamefont{Wise}},
  \bibinfo{author}{\bibfnamefont{K.}~\bibnamefont{Chatterjee}},
  \bibinfo{author}{\bibfnamefont{M.}~\bibnamefont{Yi}},
  \bibinfo{author}{\bibfnamefont{T.}~\bibnamefont{Kondo}},
  \bibinfo{author}{\bibfnamefont{T.}~\bibnamefont{Takeuchi}},
  \bibinfo{author}{\bibfnamefont{H.}~\bibnamefont{Ikuta}}, \bibnamefont{and}
  \bibinfo{author}{\bibfnamefont{E.~W.} \bibnamefont{Hudson}},
  \bibinfo{journal}{Nature Physics} \textbf{\bibinfo{volume}{3}},
  \bibinfo{pages}{802} (\bibinfo{year}{2007}).

\bibitem[{\citenamefont{Norman et~al.}(1998)\citenamefont{Norman, Ding,
  Randeria, Campuzano, Yokoya, Takeuchi, Takahashi, Mochiku, Kadowaki,
  Guptasarma et~al.}}]{FERMIARCS}
\bibinfo{author}{\bibfnamefont{M.~R.} \bibnamefont{Norman}},
  \bibinfo{author}{\bibfnamefont{H.}~\bibnamefont{Ding}},
  \bibinfo{author}{\bibfnamefont{M.}~\bibnamefont{Randeria}},
  \bibinfo{author}{\bibfnamefont{J.~C.} \bibnamefont{Campuzano}},
  \bibinfo{author}{\bibfnamefont{T.}~\bibnamefont{Yokoya}},
  \bibinfo{author}{\bibfnamefont{T.}~\bibnamefont{Takeuchi}},
  \bibinfo{author}{\bibfnamefont{T.}~\bibnamefont{Takahashi}},
  \bibinfo{author}{\bibfnamefont{T.}~\bibnamefont{Mochiku}},
  \bibinfo{author}{\bibfnamefont{K.}~\bibnamefont{Kadowaki}},
  \bibinfo{author}{\bibfnamefont{P.}~\bibnamefont{Guptasarma}},
  \bibnamefont{et~al.}, \bibinfo{journal}{Nature}
  \textbf{\bibinfo{volume}{392}}, \bibinfo{pages}{157} (\bibinfo{year}{1998}).

\bibitem[{\citenamefont{Kanigel et~al.}(2006)\citenamefont{Kanigel, Norman,
  Randeria, Chatterjee, Souma, Kaminski, Fretwell, Rosenkranz, Shi, Sato
  et~al.}}]{FERMIARCS2}
\bibinfo{author}{\bibfnamefont{A.}~\bibnamefont{Kanigel}},
  \bibinfo{author}{\bibfnamefont{M.~R.} \bibnamefont{Norman}},
  \bibinfo{author}{\bibfnamefont{M.}~\bibnamefont{Randeria}},
  \bibinfo{author}{\bibfnamefont{U.}~\bibnamefont{Chatterjee}},
  \bibinfo{author}{\bibfnamefont{S.}~\bibnamefont{Souma}},
  \bibinfo{author}{\bibfnamefont{A.}~\bibnamefont{Kaminski}},
  \bibinfo{author}{\bibfnamefont{H.~M.} \bibnamefont{Fretwell}},
  \bibinfo{author}{\bibfnamefont{S.}~\bibnamefont{Rosenkranz}},
  \bibinfo{author}{\bibfnamefont{M.}~\bibnamefont{Shi}},
  \bibinfo{author}{\bibfnamefont{T.}~\bibnamefont{Sato}}, \bibnamefont{et~al.},
  \bibinfo{journal}{Nature Physics} \textbf{\bibinfo{volume}{2}},
  \bibinfo{pages}{447} (\bibinfo{year}{2006}).

\bibitem[{\citenamefont{Valla et~al.}(2006)\citenamefont{Valla, Fedorov, Lee,
  Davis, and Gu}}]{VALLA2}
\bibinfo{author}{\bibfnamefont{T.}~\bibnamefont{Valla}},
  \bibinfo{author}{\bibfnamefont{A.~V.} \bibnamefont{Fedorov}},
  \bibinfo{author}{\bibfnamefont{J.}~\bibnamefont{Lee}},
  \bibinfo{author}{\bibfnamefont{J.~C.} \bibnamefont{Davis}}, \bibnamefont{and}
  \bibinfo{author}{\bibfnamefont{G.~D.} \bibnamefont{Gu}},
  \bibinfo{journal}{Science} \textbf{\bibinfo{volume}{22}},
  \bibinfo{pages}{1914} (\bibinfo{year}{2006}).

\bibitem[{\citenamefont{Storey et~al.}(2007)\citenamefont{Storey, Tallon,
  Williams, and Loram}}]{STOREYRAMAN}
\bibinfo{author}{\bibfnamefont{J.~G.} \bibnamefont{Storey}},
  \bibinfo{author}{\bibfnamefont{J.~L.} \bibnamefont{Tallon}},
  \bibinfo{author}{\bibfnamefont{G.~V.~M.} \bibnamefont{Williams}},
  \bibnamefont{and} \bibinfo{author}{\bibfnamefont{J.~W.} \bibnamefont{Loram}},
  \bibinfo{journal}{Phys. Rev. B} \textbf{\bibinfo{volume}{76}},
  \bibinfo{pages}{060502(R)} (\bibinfo{year}{2007}).

\bibitem[{\citenamefont{Storey et~al.}(2008{\natexlab{a}})\citenamefont{Storey,
  Tallon, and Williams}}]{STOREYRAMAN2}
\bibinfo{author}{\bibfnamefont{J.~G.} \bibnamefont{Storey}},
  \bibinfo{author}{\bibfnamefont{J.~L.} \bibnamefont{Tallon}},
  \bibnamefont{and} \bibinfo{author}{\bibfnamefont{G.~V.~M.}
  \bibnamefont{Williams}}, \bibinfo{journal}{CAP} \textbf{\bibinfo{volume}{8}},
  \bibinfo{pages}{280} (\bibinfo{year}{2008}{\natexlab{a}}).

\bibitem[{\citenamefont{Zheng et~al.}(2005)\citenamefont{Zheng, Kuhns, Reyes,
  Liang, and Lin}}]{ZHENG}
\bibinfo{author}{\bibfnamefont{G.~Q.} \bibnamefont{Zheng}},
  \bibinfo{author}{\bibfnamefont{P.~L.} \bibnamefont{Kuhns}},
  \bibinfo{author}{\bibfnamefont{A.~P.} \bibnamefont{Reyes}},
  \bibinfo{author}{\bibfnamefont{B.}~\bibnamefont{Liang}}, \bibnamefont{and}
  \bibinfo{author}{\bibfnamefont{C.~T.} \bibnamefont{Lin}},
  \bibinfo{journal}{Phys. Rev. Lett.} \textbf{\bibinfo{volume}{94}},
  \bibinfo{pages}{047006} (\bibinfo{year}{2005}).

\bibitem[{\citenamefont{Storey et~al.}(2008{\natexlab{b}})\citenamefont{Storey,
  Tallon, and Williams}}]{STOREYENTROPY}
\bibinfo{author}{\bibfnamefont{J.~G.} \bibnamefont{Storey}},
  \bibinfo{author}{\bibfnamefont{J.~L.} \bibnamefont{Tallon}},
  \bibnamefont{and} \bibinfo{author}{\bibfnamefont{G.~V.~M.}
  \bibnamefont{Williams}}, \bibinfo{journal}{Phys. Rev. B}
  \textbf{\bibinfo{volume}{77}}, \bibinfo{pages}{052504}
  (\bibinfo{year}{2008}{\natexlab{b}}).

\bibitem[{\citenamefont{Loram et~al.}(2001)\citenamefont{Loram, Luo, Cooper,
  Liang, and Tallon}}]{ENTROPYDATA2}
\bibinfo{author}{\bibfnamefont{J.~W.} \bibnamefont{Loram}},
  \bibinfo{author}{\bibfnamefont{J.}~\bibnamefont{Luo}},
  \bibinfo{author}{\bibfnamefont{J.~R.} \bibnamefont{Cooper}},
  \bibinfo{author}{\bibfnamefont{W.~Y.} \bibnamefont{Liang}}, \bibnamefont{and}
  \bibinfo{author}{\bibfnamefont{J.~L.} \bibnamefont{Tallon}},
  \bibinfo{journal}{J. Phys. Chem. Solids.} \textbf{\bibinfo{volume}{62}},
  \bibinfo{pages}{59} (\bibinfo{year}{2001}).

\bibitem[{\citenamefont{Kaminski et~al.}(2006)\citenamefont{Kaminski,
  Rosenkranz, Fretwell, Norman, Randeria, Campuzano, Park, Li, and
  Raffy}}]{2212VHS}
\bibinfo{author}{\bibfnamefont{A.}~\bibnamefont{Kaminski}},
  \bibinfo{author}{\bibfnamefont{S.}~\bibnamefont{Rosenkranz}},
  \bibinfo{author}{\bibfnamefont{H.~M.} \bibnamefont{Fretwell}},
  \bibinfo{author}{\bibfnamefont{M.~R.} \bibnamefont{Norman}},
  \bibinfo{author}{\bibfnamefont{M.}~\bibnamefont{Randeria}},
  \bibinfo{author}{\bibfnamefont{J.~C.} \bibnamefont{Campuzano}},
  \bibinfo{author}{\bibfnamefont{J.~M.} \bibnamefont{Park}},
  \bibinfo{author}{\bibfnamefont{Z.~Z.} \bibnamefont{Li}}, \bibnamefont{and}
  \bibinfo{author}{\bibfnamefont{H.}~\bibnamefont{Raffy}},
  \bibinfo{journal}{Phys. Rev. B} \textbf{\bibinfo{volume}{73}},
  \bibinfo{pages}{174511} (\bibinfo{year}{2006}).

\bibitem[{\citenamefont{Ding et~al.}(1996)\citenamefont{Ding, Bellman,
  Campuzano, Randeria, Norman, Yokoya, Takahashi, Katayama-Yoshida, Mochiku,
  Kadowaki et~al.}}]{DING}
\bibinfo{author}{\bibfnamefont{H.}~\bibnamefont{Ding}},
  \bibinfo{author}{\bibfnamefont{A.~F.} \bibnamefont{Bellman}},
  \bibinfo{author}{\bibfnamefont{J.~C.} \bibnamefont{Campuzano}},
  \bibinfo{author}{\bibfnamefont{M.}~\bibnamefont{Randeria}},
  \bibinfo{author}{\bibfnamefont{M.~R.} \bibnamefont{Norman}},
  \bibinfo{author}{\bibfnamefont{T.}~\bibnamefont{Yokoya}},
  \bibinfo{author}{\bibfnamefont{T.}~\bibnamefont{Takahashi}},
  \bibinfo{author}{\bibfnamefont{H.}~\bibnamefont{Katayama-Yoshida}},
  \bibinfo{author}{\bibfnamefont{T.}~\bibnamefont{Mochiku}},
  \bibinfo{author}{\bibfnamefont{K.}~\bibnamefont{Kadowaki}},
  \bibnamefont{et~al.}, \bibinfo{journal}{Phys. Rev. Lett}
  \textbf{\bibinfo{volume}{76}}, \bibinfo{pages}{1533} (\bibinfo{year}{1996}).

\bibitem[{\citenamefont{Kordyuk et~al.}(2003)\citenamefont{Kordyuk, Borisenko,
  Knupfer, and Fink}}]{UD77K}
\bibinfo{author}{\bibfnamefont{A.~A.} \bibnamefont{Kordyuk}},
  \bibinfo{author}{\bibfnamefont{S.~V.} \bibnamefont{Borisenko}},
  \bibinfo{author}{\bibfnamefont{M.}~\bibnamefont{Knupfer}}, \bibnamefont{and}
  \bibinfo{author}{\bibfnamefont{J.}~\bibnamefont{Fink}},
  \bibinfo{journal}{Phys. Rev. B} \textbf{\bibinfo{volume}{67}},
  \bibinfo{pages}{064504} (\bibinfo{year}{2003}).

\bibitem[{\citenamefont{Yu et~al.}(2007)\citenamefont{Yu, Munzar, Boris,
  Yordanov, Chaloupka, Wolf, Lin, Keimer, and Bernhard}}]{YU}
\bibinfo{author}{\bibfnamefont{L.}~\bibnamefont{Yu}},
  \bibinfo{author}{\bibfnamefont{D.}~\bibnamefont{Munzar}},
  \bibinfo{author}{\bibfnamefont{A.~V.} \bibnamefont{Boris}},
  \bibinfo{author}{\bibfnamefont{P.}~\bibnamefont{Yordanov}},
  \bibinfo{author}{\bibfnamefont{J.}~\bibnamefont{Chaloupka}},
  \bibinfo{author}{\bibfnamefont{T.}~\bibnamefont{Wolf}},
  \bibinfo{author}{\bibfnamefont{C.~T.} \bibnamefont{Lin}},
  \bibinfo{author}{\bibfnamefont{B.}~\bibnamefont{Keimer}}, \bibnamefont{and}
  \bibinfo{author}{\bibfnamefont{C.}~\bibnamefont{Bernhard}},
  \bibinfo{journal}{arXiv:0705.0111}  (\bibinfo{year}{2007}).

\bibitem[{\citenamefont{Doiron-Leyraud
  et~al.}(2007)\citenamefont{Doiron-Leyraud, Proust, {LeBoeuf}, Levallois,
  Bonnemaison, Liang, Bonn, Hardy, and Taillefer}}]{DOIRON}
\bibinfo{author}{\bibfnamefont{N.}~\bibnamefont{Doiron-Leyraud}},
  \bibinfo{author}{\bibfnamefont{C.}~\bibnamefont{Proust}},
  \bibinfo{author}{\bibfnamefont{D.}~\bibnamefont{{LeBoeuf}}},
  \bibinfo{author}{\bibfnamefont{J.}~\bibnamefont{Levallois}},
  \bibinfo{author}{\bibfnamefont{J.-B.} \bibnamefont{Bonnemaison}},
  \bibinfo{author}{\bibfnamefont{R.}~\bibnamefont{Liang}},
  \bibinfo{author}{\bibfnamefont{D.~A.} \bibnamefont{Bonn}},
  \bibinfo{author}{\bibfnamefont{W.~N.} \bibnamefont{Hardy}}, \bibnamefont{and}
  \bibinfo{author}{\bibfnamefont{L.}~\bibnamefont{Taillefer}},
  \bibinfo{journal}{Nature} \textbf{\bibinfo{volume}{447}},
  \bibinfo{pages}{565} (\bibinfo{year}{2007}).

\bibitem[{\citenamefont{Yelland et~al.}(2008)\citenamefont{Yelland, Singleton,
  Mielke, Harrison, Balakirev, Dabrowski, and Cooper}}]{YELLAND}
\bibinfo{author}{\bibfnamefont{E.~A.} \bibnamefont{Yelland}},
  \bibinfo{author}{\bibfnamefont{J.}~\bibnamefont{Singleton}},
  \bibinfo{author}{\bibfnamefont{C.~H.} \bibnamefont{Mielke}},
  \bibinfo{author}{\bibfnamefont{N.}~\bibnamefont{Harrison}},
  \bibinfo{author}{\bibfnamefont{F.~F.} \bibnamefont{Balakirev}},
  \bibinfo{author}{\bibfnamefont{B.}~\bibnamefont{Dabrowski}},
  \bibnamefont{and} \bibinfo{author}{\bibfnamefont{J.~R.}
  \bibnamefont{Cooper}}, \bibinfo{journal}{Phys. Rev. Lett.}
  \textbf{\bibinfo{volume}{100}}, \bibinfo{pages}{047003}
  (\bibinfo{year}{2008}).

\bibitem[{\citenamefont{Bangura et~al.}(2008)\citenamefont{Bangura, Fletcher,
  Carrington, Levallois, Nardone, Heard, Doiron-Leyraud, LeBoeuf, Taillefer,
  Adachi et~al.}}]{BANGURA}
\bibinfo{author}{\bibfnamefont{A.~F.} \bibnamefont{Bangura}},
  \bibinfo{author}{\bibfnamefont{J.~D.} \bibnamefont{Fletcher}},
  \bibinfo{author}{\bibfnamefont{A.}~\bibnamefont{Carrington}},
  \bibinfo{author}{\bibfnamefont{J.}~\bibnamefont{Levallois}},
  \bibinfo{author}{\bibfnamefont{B.}~\bibnamefont{Nardone},
  \bibfnamefont{M.~Vignolle}}, \bibinfo{author}{\bibfnamefont{P.~J.}
  \bibnamefont{Heard}},
  \bibinfo{author}{\bibfnamefont{N.}~\bibnamefont{Doiron-Leyraud}},
  \bibinfo{author}{\bibfnamefont{D.}~\bibnamefont{LeBoeuf}},
  \bibinfo{author}{\bibfnamefont{L.}~\bibnamefont{Taillefer}},
  \bibinfo{author}{\bibfnamefont{S.}~\bibnamefont{Adachi}},
  \bibnamefont{et~al.}, \bibinfo{journal}{Phys. Rev. Lett.}
  \textbf{\bibinfo{volume}{100}}, \bibinfo{pages}{047004}
  (\bibinfo{year}{2008}).

\bibitem[{\citenamefont{Kondo et~al.}(2007{\natexlab{b}})\citenamefont{Kondo,
  Khasanov, Karpinski, Kazakov, Zhigadlo, Ohta, Fretwell, Palczewski, Koll,
  Mesot et~al.}}]{KONDO124}
\bibinfo{author}{\bibfnamefont{T.}~\bibnamefont{Kondo}},
  \bibinfo{author}{\bibfnamefont{R.}~\bibnamefont{Khasanov}},
  \bibinfo{author}{\bibfnamefont{J.}~\bibnamefont{Karpinski}},
  \bibinfo{author}{\bibfnamefont{S.~M.} \bibnamefont{Kazakov}},
  \bibinfo{author}{\bibfnamefont{N.~D.} \bibnamefont{Zhigadlo}},
  \bibinfo{author}{\bibfnamefont{T.}~\bibnamefont{Ohta}},
  \bibinfo{author}{\bibfnamefont{H.~M.} \bibnamefont{Fretwell}},
  \bibinfo{author}{\bibfnamefont{A.~D.} \bibnamefont{Palczewski}},
  \bibinfo{author}{\bibfnamefont{J.~D.} \bibnamefont{Koll}},
  \bibinfo{author}{\bibfnamefont{J.}~\bibnamefont{Mesot}},
  \bibnamefont{et~al.}, \bibinfo{journal}{Phys. Rev. Lett.}
  \textbf{\bibinfo{volume}{98}}, \bibinfo{pages}{157002}
  (\bibinfo{year}{2007}{\natexlab{b}}).

\bibitem[{\citenamefont{Borisenko et~al.}(2006)\citenamefont{Borisenko,
  Kordyuk, Zabolotnyy, Geck, Inosov, Koitzsch, Fink, Knupfer, B\"{u}chner,
  Hinkov et~al.}}]{BORISENKO}
\bibinfo{author}{\bibfnamefont{S.~V.} \bibnamefont{Borisenko}},
  \bibinfo{author}{\bibfnamefont{A.~A.} \bibnamefont{Kordyuk}},
  \bibinfo{author}{\bibfnamefont{V.}~\bibnamefont{Zabolotnyy}},
  \bibinfo{author}{\bibfnamefont{J.}~\bibnamefont{Geck}},
  \bibinfo{author}{\bibfnamefont{D.}~\bibnamefont{Inosov}},
  \bibinfo{author}{\bibfnamefont{A.}~\bibnamefont{Koitzsch}},
  \bibinfo{author}{\bibfnamefont{J.}~\bibnamefont{Fink}},
  \bibinfo{author}{\bibfnamefont{M.}~\bibnamefont{Knupfer}},
  \bibinfo{author}{\bibfnamefont{B.}~\bibnamefont{B\"{u}chner}},
  \bibinfo{author}{\bibfnamefont{V.}~\bibnamefont{Hinkov}},
  \bibnamefont{et~al.}, \bibinfo{journal}{Phys. Rev. Lett.}
  \textbf{\bibinfo{volume}{96}}, \bibinfo{pages}{117004}
  (\bibinfo{year}{2006}).

\bibitem[{\citenamefont{Feng et~al.}(2001)\citenamefont{Feng, Armitage, Lu,
  Damascelli, Hu, Bogdanov, Lanzara, Ronning, Shen, Eisaki et~al.}}]{FENG}
\bibinfo{author}{\bibfnamefont{D.~L.} \bibnamefont{Feng}},
  \bibinfo{author}{\bibfnamefont{N.~P.} \bibnamefont{Armitage}},
  \bibinfo{author}{\bibfnamefont{D.~H.} \bibnamefont{Lu}},
  \bibinfo{author}{\bibfnamefont{A.}~\bibnamefont{Damascelli}},
  \bibinfo{author}{\bibfnamefont{J.~P.} \bibnamefont{Hu}},
  \bibinfo{author}{\bibfnamefont{P.}~\bibnamefont{Bogdanov}},
  \bibinfo{author}{\bibfnamefont{A.}~\bibnamefont{Lanzara}},
  \bibinfo{author}{\bibfnamefont{F.}~\bibnamefont{Ronning}},
  \bibinfo{author}{\bibfnamefont{K.~M.} \bibnamefont{Shen}},
  \bibinfo{author}{\bibfnamefont{H.}~\bibnamefont{Eisaki}},
  \bibnamefont{et~al.}, \bibinfo{journal}{Phys. Rev. Lett}
  \textbf{\bibinfo{volume}{86}}, \bibinfo{pages}{5550} (\bibinfo{year}{2001}).

\bibitem[{\citenamefont{Pringle et~al.}(2000)\citenamefont{Pringle, Williams,
  and Tallon}}]{PRINGLE}
\bibinfo{author}{\bibfnamefont{D.~J.} \bibnamefont{Pringle}},
  \bibinfo{author}{\bibfnamefont{G.~V.~M.} \bibnamefont{Williams}},
  \bibnamefont{and} \bibinfo{author}{\bibfnamefont{J.~L.}
  \bibnamefont{Tallon}}, \bibinfo{journal}{Phys. Rev. B}
  \textbf{\bibinfo{volume}{62}}, \bibinfo{pages}{12527} (\bibinfo{year}{2000}).

\bibitem[{\citenamefont{Crawford et~al.}(1990)\citenamefont{Crawford, Farneth,
  McCarron~III, Harlow, and Moudden}}]{CRAWFORD2}
\bibinfo{author}{\bibfnamefont{M.~K.} \bibnamefont{Crawford}},
  \bibinfo{author}{\bibfnamefont{W.~E.} \bibnamefont{Farneth}},
  \bibinfo{author}{\bibfnamefont{E.~M.} \bibnamefont{McCarron~III}},
  \bibinfo{author}{\bibfnamefont{R.~L.} \bibnamefont{Harlow}},
  \bibnamefont{and} \bibinfo{author}{\bibfnamefont{A.~H.}
  \bibnamefont{Moudden}}, \bibinfo{journal}{Science}
  \textbf{\bibinfo{volume}{250}}, \bibinfo{pages}{1390} (\bibinfo{year}{1990}).

\end{thebibliography}
\end{document}